\documentclass[twocolumn,pra,aps]{revtex4}
\usepackage{graphicx}
\usepackage{amsmath}
\usepackage{amssymb}
\usepackage{centernot}
\usepackage{enumerate}

\begin{document}

\title{The limits of quantum superposition:\\Should ``Schr\"{o}dinger's cat" and ``Wigner's friend" be considered ``miracle" narratives?}

\author{Antoine Suarez}
\address{Center for Quantum Philosophy \\ Ackermannstrasse 25, 8044 Z\"{u}rich, Switzerland\\
suarez@leman.ch, www.quantumphil.org}

\date{June 24, 2019}

\begin{abstract}

Physicians define death as the \emph{irreversible} breakdown of all brain-functions including brain-stem. By \emph{irreversible} they mean a damage that is beyond the human capacity to restore the patient's healthy state. In the same line I propose to complete the definition of quantum physics \cite{as19} by \emph{Principle D (Detection):} \emph{Detection outcomes (like death) are ordinarily irreversible and observer-independent}. It is then argued that this principle excludes generalization of quantum superposition to visible objects bearing observer-dependent outcomes. However this exclusion is \emph{not absolute}: It rather means that ``Schr\"{o}dinger's cat" and ``Wigner's friend" should be considered ``miracle" narratives beyond the domain of science.
\ \\


\end{abstract}


\maketitle

\section{Introduction}\label{intro}
In a previous article we have remembered how quantum physics arose from single particle interference \cite{as19}. According to the Copenhagen interpretation the moment of detection is crucial and a so called ``wavefunction collapse'' is postulated. This term conflates two different assumptions:

a) The decision of the outcome (which of the detectors counts) happens at the moment of detection.

b) At detection the outcome becomes ``irreversibly recorded" and can be observed.

It was argued \cite{as19} that assumption a) implies nonlocal coordination of decision events at detectors, and the main interpretations of quantum (Copenhagen, de Broglie-Bohm, Many-Worlds) more or less explicitly endorse the following two Principles:

\begin{itemize}
  \item \emph{Principle A (Accessibility):} \emph{All that is in space-time is accessible to observation (except in case of space-like separation).}
  \item \emph{Principle Q (Quantum): Not all what matters for physical phenomena is contained in space-time.}
\end{itemize}

In this letter we will focus on assumption b) and propose to complete the \emph{Principles A and Q} with a third principle:

\begin{itemize}
  \item \emph{Principle D (Detection): Detection outcomes (like death) are} ordinarily \emph{irreversible and observer-independent.}
\end{itemize}

In the coming Section \ref{outcome} we work out accurately the meaning of ``detection outcome" in \emph{Principle D}. In Sections \ref{born}-\ref{fr} we show that ``Schr\"{o}dinger's cat" and ``Wigner's friend" amount to assume observer-dependent outcomes and therefore cannot be considered ordinary phenomena. In Section \ref{wigner} we argue that ``Schr\"{o}dinger's cat" and ``Wigner's friend" should be considered ``miracle" narratives rather than scientific descriptions, i.e.: ``quantum phenomena" beyond quantum physics. In Sections \ref{measurement}-\ref{concluding} we draw some conclusions.

\section{``Outcomes" are \emph{ordinarily} irreversible and observer-independent}\label{outcome}
Quantum experiments are characterised by the fact that in each single round \emph{one and only one} of different possible alternative outcomes happens. So, to have quantum experiments the setup must contain at least 2 detectors (``two state system" described by ``Hilbert-space with dimension 2"). ``Born's rule" assigns probabilities to possible alternative detection outcomes. Hence, quantum physics is based on the concept of ``outcome" or ``experimental result", and for Born's rule to make sense, one has to define unambiguously what the term \emph{detection outcome} means.

The fact that at ``detection" something ``irreversible" happens has been emphasized by John A. Wheeler (also as ``Bohr's point"): ``No elementary quantum phenomenon is a phenomenon until it is a registered (`observed', `indelibly recorded') phenomenon, `brought to a close' by `an irreversible act of amplification'." \cite{jw}.

Niels Bohr on his turn postulated the ``necessity of discriminating in each experimental arrangement between those parts of the physical system considered which are to be treated as measuring instruments and those which constitute the objects under investigation" and referred to this necessity as ``\emph{a principal distinction between classical and quantum mechanical description of physical phenomena}" (italics by Bohr) \cite{nb}. Admittedly, Bohr seems to weaken this distinction by stating that: ``It is true that the place within each measuring device where this discrimination is made is in both cases [classical and quantum] largely a matter of convenience." But he insists that the distinction is of ``fundamental importance" in quantum theory: The measuring device has to be classical because of ``the indispensable use of classical concepts in the interpretation of all proper measurements, even though the classical theories do not suffice in accounting for the new types of regularities, with which we are concerned in atomic [quantum] physics."\cite{nb} Hence, Bohr clearly states that we cannot have quantum physics without some classical basic concepts, and this entails definite conditions defining when a result appears and can be observed, even if for the time being we don't know which these conditions are, where the transition point lies at which ``quantum possibilities" turn into ``classical certainties".

That this is the way Bohr thought, is further supported by John Bell in his famous article ``Against 'measurement'" \cite{jb90}. John refers to the book \emph{Quantum Mechanics} by L. D. Landau and E. M. Lifshitz (LL) as ``the nearest to Bohr that we have", and states: ``With Bohr [LL] insist again in the inhumanity of it all": LL emphasise that ``in speaking of `performing a measurement', we refer to the interaction of an electron with a classical `apparatus', which in no way presupposes the presence of an external observer." (see \cite{jb90} p. 35).

In this line of thinking we assume that at detection new information appears in space-time in form of some observable mark or sign (blackening, scintillation, sound) we can perceive with our senses: ``Observation is an elementary act of creation", in Wheeler's wording \cite{jw}. Once registered (``created"), this mark evolves visibly thereafter following a deterministic world-line (according to the equations of General Relativity) so that the future is \emph{completely} determined by the past, i.e.: derives \emph{necessarily} (with probability 1) from the content in the past light-cone. For instance, once the first bubble of the bubble trail of an electron in a liquid hydrogen chamber does appear, successive plots will deterministically follow in natural units of time.

But one could still ask: When \emph{precisely} does the wavefunction collapse happen and the first bubble appear? The same question can be put with relation to the detection of a photon by a photomultiplier: When \emph{exactly} after the wave's arrival can we say the detection takes place? When does a process of amplification become \emph{irreversible} and produce a registered phenomenon, a detected outcome? \cite{jw}.

In this context John Bell liked to invoke GRW's ``spontaneous wave function collapse'' \cite{bg, jb90}. Later Roger Penrose introduced ``gravitationally induced decoherence'' \cite{rp}. These processes are also denoted ``objective reduction''(OR) \cite{rp} because they dispose explicitly of the assumption that a human observer has to be \emph{actually} present for a registration to take place.

But it may also be profitable to consider a daily process that is generally considered to be irreversible in principle: I mean \emph{death}. The medical definition of death includes explicitly the concept of \emph{irreversibility} as it basically defines death as the ``irreversible" break-down of all the brain functions included brainstem. More precisely physicians declare someone dead after establishing the so called ``clinical signs of death" by checking the absence of certain spontaneous movements, especially spontaneous breathing. But what do they mean by \emph{irreversible}? Just that a damage happens beyond our capabilities to repair. In establishing death this way, we are assuming as obvious that our capacity of restoring neuronal dynamics (our repairing capability) is limited in principle, even if we don't yet know where this limitation comes from. For someone to die it is not necessary that a physician is watching him, and to this extent dead is ``observer-independent"; however it is \emph{humanity-dependent}, in the sense that it depends on the human capabilities to ``repair diseases".

We propose to consider detection \emph{irreversible} in the same sense physicians consider dead \emph{irreversible}: At detection something happens beyond our capabilities to restore. So for instance a process of amplification in a photomultiplier becomes ``irreversible in principle" at a certain level, if as soon as this level is reached an operation beyond the human capabilities would be required to restore the quantum state of the incoming photon. When such a level is reached the detector counts.

Such a view combines the \emph{subjective} and \emph{objective} interpretation of measurement: On the one hand no human observer has to be actually present in order that a registration takes place, just the same as in the GRW's ``spontaneous collapse'' \cite{bg}) or Penrose's ``objective reduction'' (OR) \cite{rp}; on the other hand the process by which an outcome becomes registered (the ``collapse" or ``reduction") is defined with relation to the capabilities of human observers (the way the human brain functions after all). Detection is \emph{observer-independent} but \emph{humanity-dependent}: A process is \emph{irreversible} because the complexity of the reversing task is beyond human operational capabilities. For the time being the conditions defining this \emph{irreversibility} threshold (likely involving a new constant of nature) are unknown: This is the ``measurement problem" (see Section \ref{measurement}).

So, if a system collapses, it collapses for all observers; performed experiments have \emph{observer-independent} results.

Accordingly, the term \emph{outcome} in \emph{Principle D} (Section \ref{intro}) means the \emph{particular} detection event (among different alternative ones) that comes to happen in a certain experimental round; outcomes are \emph{ordinarily} irreversible and observer-independent.

We introduce the term \emph{`ordinarily'} in this definition because regarding \emph{irreversibility} we share the Fuchs-Peres' view that ``it is a practical problem, not a fundamental one." \cite{fp} We think that detection like dead are \emph{not absolutely} irreversible (nothing speaks for instance against believing in the resurrection of Jesus Christ) but only \emph{ordinarily}, in the sense that both processes lay beyond our capabilities to restore the initial state. Obviously if by ``practical problems" one means such ``that presently are unsolved but one day we will be able to overcome", then detection and dead are not only practical problems but fundamental ones: For reversing them we ``would need \emph{complete control} of all microscopic degrees of freedom" \cite{fp}, and this is something we will \emph{never} achieve, as we will never achieve phoning faster than light or microscopes with arbitrary high resolution (See Section \ref{measurement}).

\begin{figure}
\includegraphics[trim = 2mm 85mm 10mm 0mm, clip, width=0.99\columnwidth]{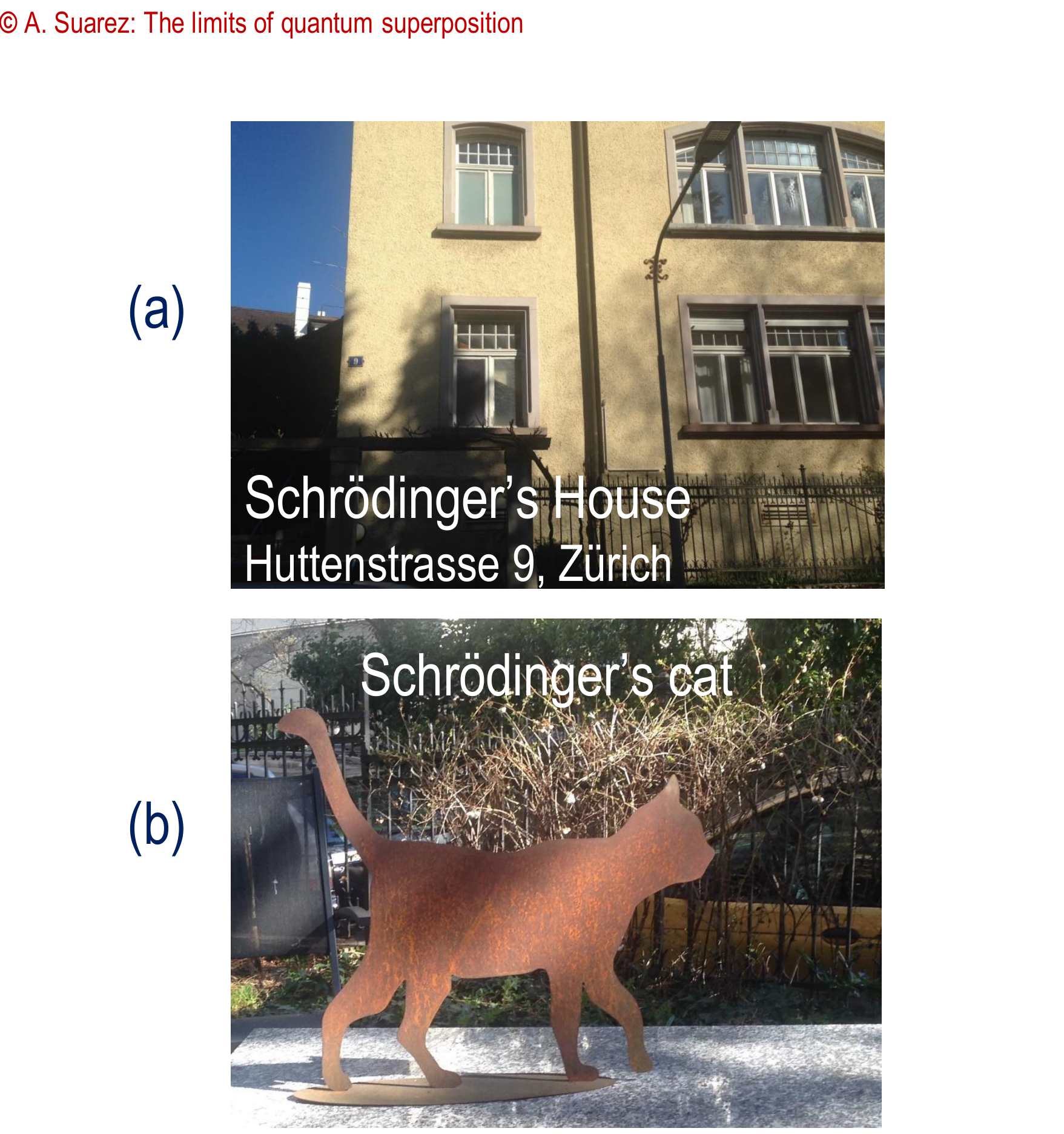}
\caption{(a) House where Erwin Schr\"{o}dinger lived while he sojourned in Z\"{u}rich. (b) Sheet metal cat puppet the house owner has put in the garden as a memorial of Schr\"{o}dinger's ``cat paradox": The cat is not fixed on the ground, so the visitor cannot know in advance with certainty where he will meet the cat. If you had visited the house immediately after I was there, you would have met the cat on the garden table I put it to take the picture in (b).
According to ``Schr\"{o}dinger's cat" assumption, a cat can be considered as being in superposition of two alternative states `alive' and `dead'. But then nothing speaks against considering  Schr\"{o}dinger's metal cat in (b) in superposition of different alternative locations in the garden of the house at (a) (see Section \ref{born}).}
\label{cat}
\end{figure}

\section{``Born's rule" is \emph{ordinarily} at variance with ``Schr\"{o}dinger's cat" and ``Wigner's friend"}\label{born}
It is well known that Erwin Schr\"{o}dinger with his famous ``cat paradox" has popularized an interpretation of quantum theory,``which does not impose any constraints on the complexity of objects it is applied to" \cite{fr}. Schr\"{o}dinger's gedanken-experiment generalizes quantum superposition to visible objects: A cat can be considered as being in superposition of two alternative states `alive' and `dead'. But then nothing speaks against considering  Schr\"{o}dinger's metal cat in Z\"{u}rich in superposition of different alternative locations in the garden of the house at the Huttenstrasse 9 (Fig.\ref{cat}). That is, the metal cat puppet would not be compelled to follow a deterministic world-line but could move whirling around in the garden, or apparating and dissaparating at different locations.

On the other hand the ``cat paradox" assumes that an observer can apply Born's rule to arbitrary systems, including large ones that may contain other observers; the system can even be an entire lab \cite{fr}. This means that the moment when a system collapses depends on the particular observer watching it; a system can collapse for certain observers and not for others: physical reality becomes \emph{radically} observer-dependent. This consequence comes into full effect in ``Wigner's Freund" gedanken-experiment (\cite{ew}, p. 179-181):

Suppose an experiment in lab L, where Wigner's friend (F) using Born's rule predicts outcome 0 (detector A counts) with probability $1/2$ and outcome 1 (detector B counts) with probability $1/2$, as for instance the experiment in \cite{Guerreiro12}. Suppose further that in a particular round of the experiment F performs a measurement and observes actually outcome 0 at time $t$.

By contrast, if W applies Born's rule to the whole lab L, W must conclude that F at the same time $t$ has neither observed 0 nor 1.

This means that if W applies Born's rule to the whole lab L, he is led to a prediction about what F observes, which is different from what F actually observes. Eugen P. Wigner concluded that to avoid the paradox one should assume that the wave-function was already collapsed by F at the very moment he becomes aware of the outcome, and so at this same moment it was also collapsed for W. It is well known that the weakness of this solution is the assumption that the collapse depends on ``consciousness" and hence is intrinsically \emph{subjective}: W doesn't know whether or not F was aware at time $t$, and could in principle very well be aware before F, so that nothing forbids W to apply Born's rule to L.

This problem disappears if one applies the definition of \emph{outcome} according to the preceding Section \ref{outcome}:
Wigner (W) outside the lab must conclude that at time $t$ in L \emph{either} outcome 0 (a count in detector A) \emph{or} the opposite outcome 1 (a count in detector B) was \emph{irreversibly} registered. Accordingly the wave-function collapsed at time $t$ no matter whether or not F was aware of it.

According to \emph{Principle D}, the outcome F observes and that W predicts \emph{ordinarily} overlap. Quantum superposition is limited by \emph{irreversibility} of detection: The very presence of the term `outcome' in Borns's rule entails that the rule does not apply to visible objects, unless in case of some \emph{extraordinary} phenomenon, that is, a phenomenon where ``ireversibility" becomes reversed beyond human operational capabilities. And the other way around: The very assumption of ``Schr\"{o}dinger's cat" entails to give up \emph{certainty} about the existence of definite states of visible bodies (either `cat alive' or `cat dead') at a determined time (immediately after detection), and therefore, with more reason, also the validity of predictions with \emph{certainty} according to Born's rule.

It is really noteworthy that in order to keep Born's rule also in case of predictions with \emph{certainty} and observer-independent outcomes, one has to limit quantum superposition and avoid its application to visible systems like cats, and human experimenters.

\section{Frauchiger-Renner's theorem: ``Wigner's friend" as ``super-natural" phenomenon}\label{fr}
The conclusion of the preceding Section \ref{born} is strengthened by Frauchiger-Renner's no-go theorem \cite{fr}. The theorem uses an improved version of ``Wigner's friend" gedanken-experiment and introduces three Assumptions:

\begin{itemize}
  \item (\textsf{Q}) \emph{Universal validity of quantum theory}: It must be possible to employ together (1) quantum theory to model complex systems that include agents who are themselves using quantum theory, and (2) Born's rule in case of predictions with \emph{certainty}.
  \item (\textsf{C}) \emph{Consistency}: Different agents’ predictions are not contradictory.
  \item (\textsf{S}) \emph{Single outcome}: From the viewpoint of an agent who carries out a particular measurement, this measurement has one single outcome.
\end{itemize}

It is then proved that these three Assumptions cannot hold together: By extending quantum superposition to visible objects \emph{and} accepting predictions with \emph{certainty} of Born's rule (\textsf{Q}), one is led to deny (\textsf{C}) inferring that physical reality is \emph{unrestrictedly} ``observer-dependent" (one agent, upon observing a particular measurement outcome, must conclude that another agent has predicted the opposite outcome with certainty), or deny (\textsf{S}) inferring violation of conservation of energy (in the experiment of \cite{Guerreiro12} one photon would produce two counts). And viceversa, if one keeps to (\textsf{C}) and/or (\textsf{S}), one has to give up (\textsf{Q}) i.e.: either quantum superposition of visible objects or Born's rule predictions with \emph{certainty}.

Interestingly the no-go theorem \cite{fr} does not care about ``irreversibility". To this extent the introduction of (\textsf{Q}) rests on an equivocal use of the term ``outcome" and overlooks that Born's rule by definition excludes its application to already collapsed systems (as discussed in the preceding Section \ref{born}). Admittedly, this a widespread mistake: ``Nothing in principle prevents us from quantizing a colleague" \cite{fp}. So the merit of \cite{fr} consists in showing that, if one makes this mistake, one is led to deny (\textsf{C}) or (\textsf{S}).

Does this mean that \cite{fr} proves ``quantum theory inconsistent"? Only as far as ``consistency" remains limited to the meaning conveyed in Assumption (\textsf{C}): The theorem certainly proves that the \emph{generalization} of quantum superposition to complex systems is at odds with assuming that the physical reality is ``observer-independent". But this does not mean that the \emph{generalisation} is absurd and can never take place. It is noteworthy that Frauchiger-Renner consider Assumptions (\textsf{C}) and (\textsf{S}) to be ``natural" ones. So, what the theorem in fact proves is that quantum superposition of visible objects and violation of conservation of energy should not be considered ``natural" phenomena but rather ``super-natural" ones.

Assumption (\textsf{C}) is actually sort of ``probability 1" belief (in the sense of Qbism) on the basis of our \emph{ordinary} daily experience. The only thing we can ``consistently" assert is that an observer-independent ``physical reality" is highly convenient \emph{for all practical purposes}, otherwise it would be highly complicated for humanity to live and communicate conveniently.

Accordingly, \cite{fr} only excludes quantum superposition \emph{within} the limit of \emph{what is possible and useful} for human agents. Beyond this limit quantum superposition can very well happen, but bears ``super-natural" phenomena.

So with \cite{fr} one is led to a similar conclusion as with our \emph{Principle D}. In fact this Principle encompass the Assumptions (\textsf{C}) and (\textsf{S}) in \cite{fr}: The phenomena Frauchiger-Renner call ``natural" we call ``ordinary". Our concept of ``nature" includes ``extraordinary" phenomena, which from Frauchiger-Renner's point of view are ``super-natural" ones.

Frauchiger-Renner's no-go theorem has been echoed by numerous criticisms that in fact amount to rule out ``Schr\"{o}dinger's cat" and ``Wigner's friend":

So, for instance, Scot Aaronson refutes the theorem by referring to Asher Peres celebrated dictum: ``unperformed measurements have no results" (actually Peres says ``experiments" instead of ``measurements") \cite{sa18}. In our view one should rather state that ``performed experiments \emph{ordinarily} have definite \emph{irreversibly} registered results, which are \emph{observer-independent}". But this amounts to exclude ``Schr\"{o}dinger's cat" and ``Wigner's friend" at least from the realm of \emph{ordinary} physical phenomena.

An interesting criticism is provided in \cite{aas}, where it is argued that: ``information from direct measurement must trump inference from steering. The erroneous belief that both paths should lead to identical conclusions can be traced to the usual prejudice that measurements should reveal a pre-existing state of affairs." This amount to say that performed experiments have definite registered results that invalidate any inference denying them, even if it comes from applying quantum superposition. And this again means: \emph{either} to reject instantiations of Quantum as ``Wigner's friend" or Born's rule in case of predictions with \emph{certainty} (giving up Assumption (\textsf{Q}) in \cite{fr}),
\emph{or} leave the realm of \emph{ordinary} physical phenomena with observer-independent outcomes (giving up Assumption (\textsf{C}) in \cite{fr}).

For sake of clarity we use the term `quantum' to refer to all instantiations of superposition, and `quantum physics or theory' when speaking about descriptions respecting \emph{Principle D}. If one makes this distinction, then ``quantum theory consistently describes the use of itself".

\section{``Schr\"{o}dinger's cat" and ``Wigner's friend": Scientific descriptions or miracle~narratives?}\label{wigner}
The idea that the physical evidence may be observer-dependent is well illustrated by the reported event called ``Fatima's miracle of the Sun": On October 13, 2017, about 70,000 pilgrims gathered in Cova da Iria (Fatima, Portugal) perceived the Sun dancing at 2 pm. By contrast 2 billion people in the rest of the world perceived the Sun following its usual trajectory \cite{wp}. So during about 10 minutes two different groups of observers had different evidence of the ``same" physical event depending on their location. What does this mean? Since the physical reality is defined by the observations, one must conclude that what watched the 70,000 in Cova da Iria was as real (or as virtual) as what watched the two billion in the rest of the world. At the end of the ``miracle" the people gathered in Cova da Iria entered again the ordinary world and perceived the Sun following the usual trajectory.

We have here the situation where different observers get different outcomes of the same physical event (``experiment"), and some of them see the Sun deviating from a deterministic trajectory, behaving like a system in superposition of different locations. The situation is similar to that resulting from the application of quantum superposition to visible systems. In this sense, Fatima's miracle can be considered a realisation of ``Wigner's friend" or ``Schr\"{o}dinger's cat" gedanken-experiment (Section \ref{born}, Figure \ref{cat}).

\emph{Ordinarily} it holds that performed experiments have the same results for different observers (Assumption (\textsf{C}) in \cite{fr}) and quantum superposition is limited. However the omniscient mind (the ``prophet" in Specker's parable \cite{Specker60, as17}) can conceive stories with such a superposition, and therefore even produce that observers at a certain place perceive the sun spinning while other observers in another place perceive it as usual (in accord with the result in \cite{fr}). Such stories lay beyond what experimental science is all about. Most importantly: Miracles by definition never altere the ordinary regularities humans can use to predict and master physical effects and develop technologies. Consequently ``miraculous events" cannot be proved to have happened on the basis of the \emph{generally} accessible ordinary evidence, but are rather believed because of reports by trustworthy witnesses, who had extraordinary evidence of them: Memories of such events do persist in history trough documented narratives of these witnesses, and not through archeological or geological evidence.

In summary: ``Schr\"{o}dinger's cat" and ``Wigner's friend" illustrate magnificently how quantum superposition can easily go over into descriptions that actually amount to miracle narratives close to that of Fatima's dancing Sun.

\section{For `Measurement'}\label{measurement}
John Bell's main charge against introducing `measurement' in the fundamental axioms of quantum mechanics is that ``it anchors there the shifty split of the world into `system' and `apparatus'." \cite{jb90}. This objection seems refuted by the arguments in the previous Sections. The transition point between `quantum' and `classical' becomes \emph{sharp} if it is defined with relation to the human capabilities according to \emph{Principle D} in Sections \ref{intro}-\ref{outcome}. Nonetheless ``the measurement problem'', i.e. where must we precisely draw the line between the two realms, remains a mystery still to elucidate in today's physics.

The ``measurement problem" is somewhat the physical correlate of the ``halting problem" in arithmetics. However in arithmetics we can sharply prove that at any time T there will be questions about numbers that we cannot answer with the methods available at time T (the well known Turing's theorem). By contrast in physics we cannot yet sharply prove (we only intuitively feel) that at any time T there will be physical processes (for instance death) that lie beyond our capabilities to restore, and therefore are irreversible \emph{with relation to the human capabilities}. In this sense interpretations of quantum mechanics can basically be split into two types: those acknowledging ``irreversibility" (with relation to human capabilities) and in particular death as a main constituent of physical reality, and those assuming that humans will once overcome death at will by technical means.

``Irreversibility" is the core of the ``measurement problem" and, as we have seen, has the noteworthy implication that we cannot apply quantum mechanical superposition to visible objects: Single-particle interference means that the two detectors watching the output ports of a Mach-Zehnder interferometer can be considered ``nonlocally" coordinated; by contrast it does not mean that a whole detector is in superposition of two distant locations, and apparates and dissaparates at one or the other location to produce the two different outcomes. ``Irreversibility" at detection \emph{by definition} excludes ``Schr\"{o}dinger's cat" and ``Wigner's friend" as \emph{ordinary} phenomena susceptible of scientific explanation, and keeping to both in introducing quantum theory is confusing: ``W [Wigner] should not use the Schr\"{o}dinger Equation to describe the
evolution of the state of L [the whole Lab including Wigner's friend F]!" \cite{jf}

``Schr\"{o}dinger's cat" is \emph{ordinarily} either alive or ``\emph{irreversibly} brain-dead".

This does not mean that a dead-person cannot resurrect, I dare to insist. It means only that resurrection is beyond what is possible for humans to do.

Notice that even from the perspective of ``Boltzman's second law" resurrection should be considered not impossible but only a highly improvable event. However the law should be reformulated to account for the limit of our capacity to repair, possibly by introducing a constant referring to the ``\emph{irreversibility} threshold". Paraphrasing Stefan Wolf: The key to thermodynamics may lie within the quantum measurement process.\cite{Wolf17}

The view proposed here implies that there cannot be such a thing as a ``wave function of the whole universe'', which would also include all human observers. If there is no human observer outside the wavefunction, there is no wavefunction at all. The fundamental importance of the human observer in Quantum Mechanics is the obvious consequence of the fundamental importance of observation and evidence in science: No science without observation, and no observation without observer. What Quantum seems to tell us after all is that the world is a dialogue between mighty ``non-neuronal intellects"  \cite{as07} and human ones. Only someone mad keeps speaking if there is nobody to hear at him, and the ``omniscient mind" is not mad!

Measurement (an act of ``observer participancy" \cite{jw}) links the quantum content of the ``omniscient mind" \cite{Specker60, as17} to ordinary human histories.

\section{``Wigner's Brain": Sleep~and~purposeful action}\label{sleep}
After stating that Wigner is not entitled to apply quantum superposition to his friend unless in case of ``miracle", it is important to clarify that Wigner, as every embodied human spirit, has to set his/her own brain in quantum superposition to think, decide, and behave. Certainly ``It’s hard to think when someone Hadamards your brain" \cite{sa18}. But to deliberate and decide, \emph{you} have to ``Hadamard" \emph{your own} brain:

While writing this article, my brain can be compared to a quantum interferometer, where the physiological parameters determine the optical path-length difference, and each act of typing corresponds to the outcome of an experimental round \cite{as19}. Before each act of typing the state of my brain is described by a quantum superposition of all the possible characters on my key-board. By deciding to type a particular character I collapse this superposition to produce a single outcome. Thereby I mentally steer \emph{from outside space-time} my brain's outcomes, that is, the sequences of bits they consist in. For this reason I can claim to be the author of the text I am writing: Notwithstanding time passes \cite{ng15}, my personal identity remains conserved because it is anchored outside time, it does not reduce to flow of time.

While typing I can make that the distribution of bits during a short period deviates from the quantum mechanical predictions for a large number of outcomes. By contrast these predictions could be considered valid for an observer watching me while I am sleeping. The bit-string outcome of a human brain tends towards its most probable sequence, i.e.: a meaningless bit string, in absence of purposeful control (in absence of conscious free-will), that is, during sleep. In presence of purposeful control (awake period), the bit-string outcome human brains ``print" may very well be meaningful.

\emph{Embodied consciousness} means consciousness limited by sleep. Born's rule fits well the outcomes of ``sleeping" systems acting aimlessly according to a ``random" pattern. Nonetheless it may fail to fit the outcomes of minds deviating \emph{purposefully} from random during short periods.

``Free-will comes first in the logical order" and ``is a prerequisite for understanding and for science" \cite{ng15}. Quantum physics provides the congenial environment where \emph{embodied} human consciousness and freedom can flourish.

\section{``Inexorable Laws of Nature" or Regularities to make a world ``fit~for~human~habitation"\cite{kelvin}?}\label{resur}

A further conclusion we can draw from the preceding Sections is that there are no ``Inexorable Laws of Nature":``the determinism of classical physics turns out to be an illusion, created by overrating mathematico-logical concepts. It is an idol, not an ideal in scientific research" \cite{mb54}. There is no ``Theory of Every Thing" that fits \emph{all possible} phenomena that may happen in the world \cite{caf10}; no equation or rule whatsoever can fit the \emph{whole set} of outcomes the ``omniscient mind" assigns to any group of observers.

This mind is kind to us and distributes the \emph{subset} of outcomes we \emph{usually} observe according to regularities we can grasp; the algorithms and equations we develop to calculate the world are like extensions of the genetic evolutionary algorithms allowing us to efficiently use our hands and legs to work and behave:``The agent and his devices are of one flesh." (\cite{caf17} pp. 29-30). What the so called ``physical reality" and ``laws of nature" basically define is a world ``fit for human habitation" \cite{kelvin}. With Eugen P. Wigner one can state that the reality of physical concepts (as for instance that of magnetic field) is ``synonymous with the usefulness of the concept, both for our own thinking, and for communicating with the others" (\cite{ew}, p. 188). The amazing fact of natural regularities creates an environment that facilitates our learning and enhance well-adapted behavioural patterns.

The mathematical equations we use to describe and predict the world (General Relativity, Born's rule) fit well the \emph{ordinary} regularities relevant to our life. In this sense such equations are basic ingredients of physical \emph{reality} and can be considered \emph{objective}.

But the omniscient mind may also assign outcomes deviating from these regularities under extraordinary circumstances. The times when these deviations come to happen are unpredictable in principle, the same way as the single outcomes in most quantum experiments are unpredictable. How low their \emph{absolute probability} is, we don't know, and this data may even be beyond what we will ever know. In any case the occurrence of such \emph{highly} improbable events does not break any ``inexorable law of nature" but only the limit of \emph{what is possible} for human observers. We cannot contrive nature to produce such deviations the same way we contrive it to make a friend's mobile ring by sending him a message. Producing such extraordinary events is beyond our technological capabilities, as it is phoning faster than light.

Nor can we invoke quantum mechanics to master so called ``para-normal" phenomena. It is claimed for instance that ``psi-subjects" can mentally affect a physical system directly without the mediation of classical local causes (psychokinesis), as let tables dance at a distance, or change the counting rate of a detector in an interference experiment by acting ``mentally" upon the detector without physically changing the phase-parameters. Such phenomena can be considered instantiations of ``Schr\"{o}dinger's cat". As such they are not impossible, but they lay beyond our technological capabilities: I cannot act upon devices in the lab the same way as I act upon my brain. Quantum mechanics does not forbid ``para-normal" effects, but it does not provide methods to take hold of them: Such effects are ``Schr\"{o}dinger's cats" belonging to the realm of ``non profit phenomena".

From this perspective, both the Quantum Born's rule and the equations of General Relativity, can be considered ``\emph{subjective beliefs}" we form on the basis of our ordinary daily experience, very much in agreement with what we are taught by QBism. The statement: ``The Sun will be tomorrow, at noon in Zürich, in the position \emph{X,Y,Z}" means that we would pay no price to enter a bet with payoff 1 if the sun begins to dance in the sky tomorrow at noon, and no payoff if it follows its usual trajectory. But it does not mean it is impossible for the Sun to dance tomorrow at noon.

One should distinguish two types of ``exceptions" to the established theories:

\begin{itemize}
  \item Phenomena we cannot predict because our equations are not yet good enough to describe them, as for instance Mercury’s perihelion before Einstein discovered General Relativity.
  \item Phenomena that are unpredictable in principle because they do not fit to any mathematical description, as for instance the Sun dancing in the Sky in Fatima at 2 pm on October 13, 1917.
\end{itemize}

Correspondingly, one should distinguish two types of visible phenomena: Those we can perform by physical operations (as for instance letting a mobile ring, influencing detection rates by changing phases), and those beyond our physical capabilities (letting a dead resurrect, reversing a detection, changing the counting rate of a detector through ``mental power").

The distinction between \emph{what is and is not possible} for human experimenters \cite{deutsch} is likely more important than the distinction between \emph{``classical and quantum"}.

\section{Concluding remarks}\label{concluding}
``Schr\"{o}din\-ger's cat" and ``Wigner's friend" highlight how Quantum spreads beyond the borders of the \emph{ordinary} world (where experiments have observer-independent outcomes) to the realm of \emph{extraordinary} or \emph{super-natural} phenomena (where physical reality may be observer-dependent).

The set of outcome assignments in the omniscient mind consists of two subsets: one containing the assignments defining the usual regularities we observe, and the other shaping unusual phenomena (like ``Schr\"{o}din\-ger's cat", ``Wigner's friend", Fatima's miracle of the sun):

On the one hand, the scientific rules and equations we employ to describe the world fit well \emph{a highly significant part} of physical reality: the \emph{subset} of outcomes shaping the regularities we are used to. Thereby they make it possible for us to predict, develop technologies, and live, and in this sense can be considered an \emph{``objective"} ingredient of the physical reality. What we call ``laws of nature" are simply rules the omniscient mind imposes upon the world in order we can live comfortably in, rules good for all practical purposes. This may explain the ``unreasonable effectiveness of mathematics in the natural sciences" \cite{ew60}.

On the other hand, all our rules and algorithms do not exhaust the \emph{whole} physical reality. That the sun starts spinning in the sky, or a dead resurrects are not impossible events but only \emph{highly} improbable ones. In other words, they happen with probability 0 in the \emph{ordinary} world we are used to, but not \emph{absolutely}. Similarly, the prediction that a measurement upon a determined quantum system (e.g.: a Mach-Zehnder interferometer with equal arm-lengths) yields an outcome with ``probability 1" does not mean that ``nature is obliged" to produce this outcome. In Chris Fuchs' wording: ``All probabilities, including all quantum probabilities, are so subjective they never tell nature what to do. This includes probability-1 assignments."\cite{caf} The use of probability in physics is always an ``idealization" \cite{sk15} ``for all practical purposes".

To have a world good for all practical human purposes, \emph{quantum superposition is limited by measurement}. Measurement (choices of experimenters after all) links the content of the ``omniscient mind" (``All possible worlds or histories") to what is going on in the ordinary human world. The measuring devices (detectors) are excluded from quantum superposition when they are heavy enough and capable of obtaining an \emph{irreversible} mark upon them (Bohr, \cite{caf17} p.29). Measurement produces always at least two nonlocal correlated decisions at detection, and once an outcome is ``irreversibly registered" (detected) and becomes visible, it evolves according to a relativistic local world-line: Quantum nonlocality reveals that relativistic local causality is only an appearance. Measurement lets space-time emerge from outside space-time. But if the space-time only starts with our measurements, is then the Big Bang here? To such a question John A. Wheeler answered once: ``I can imagine that we will someday have to answer your question with a `yes'." (\cite{caf}, p. 6, note 5). Without ``human free choices", no space-time!

Wigner can very well set his own brain into a quantum superposition state, but he cannot force his friend's brain to ``uncollapse" an ``irreversibly collapsed wavefunction". ``Schr\"{o}din\-ger's cat" is \emph{either} alive \emph{or} irreversibly dead. However \emph{``irreversibility"} does not mean ``impossible to reverse absolutely" but impossible to reverse by the human subject (observer and agent). When ``irreversible processes" become reversed and phenomena spontaneously deviate from the trajectories that we are used to, people of all times tends to refer to them as ``miracles": In this sense a ``miracle" does not violate any ``inexorable law of nature" but only ``rules for human convenience". Miracles help us to understand by contrast how kind the omniscient mind is to us by using such rules to make the world. The big wonder in Fatima was not that what watched the ten of thousands gathered at Cova da Iria, the dance of the Sun, but that what the 2 billion people outside observed, the Sun following its usual trajectory in the Sky, the wonder of ordinary life.

\end{document}